\documentstyle[11pt,newpasp,twoside,epsf]{article} \def\edcomment#1{\iffalse\marginpar{\raggedright\sl#1\/}\else\relax\fi}
\reversemarginpar

\begin{document} 

\title{Orbital and stellar parameters of symbiotic stars} 

 \author{Joanna Miko{\l}ajewska}

\affil{N. Copernicus Astronomical Center, Warsaw, Poland, email: 
mikolaj@camk.edu.pl}

\begin{abstract} {This paper reviews current knowledge of symbiotic binaries, with special 
emphasis on their multifrequency observational characteristics, and basic parameters of 
the symbiotic system components. We start with a brief presentation of variable phenomena 
found in symbiotic stars. This is followed by a summary of the recent progress in 
determination of their orbital and stellar parameters. We also discuss basic properties of 
the symbiotic giants compared to single evolved giants as well as the nature of the hot 
component and its outburst evolution.} \end{abstract} 

\section{Introduction} 

Symbiotic stars were isolated as a separate class of spectroscopically peculiar stars by 
Annie J. Cannon, during her work on HD Catalog,  in the beginning of XXth century. A 
typical symbiotic spectrum should display (Fig. 1) absorption features -- such as TiO 
bands and neutral metals, and an associated red continuum typical for red giants, a blue 
continuum with the H\,{\sc i} Balmer lines and jump in emission, and strong emission lines 
from relatively highly ionized species -- He\,{\sc i}, He\,{\sc ii}, [O\,{\sc iii}], etc., 
usually  found in planetary nebulae. 
The simultaneous presence in a single object of low-temperature absorption features and 
emission lines that require high excitation conditions 
apparently points to their binary nature.
In fact, binary models for symbiotic 
stars, in which a rather normal M giant was accompanied by a hot component resembling the 
central star of a planetary nebula, were proposed and discussed soon after their discovery 
(e.g. Berman 1932). Berman also proposed that the observed eruptive behaviour of some 
objects could be due to some kind of the hot component instability, perhaps similar to 
that responsible for a nova phenomenon. It is really impressive how close were these 
first models to the generally accepted present-day model.

According to this model, a typical symbiotic binary consists of an M-type giant 
transferring material to a much hotter, white dwarf companion via a stellar wind. The wind 
is ionized by the hot component giving rise to symbiotic nebula. In some systems the red 
giant is replaced by a yellow G$-$K-type giant, and the white dwarf by a main-sequence or 
neutron star. Most symbiotic stars ($\sim 80\,\%$) contain a normal giant and their near- 
IR colours show the presence of stellar photosphere, $T_{\rm eff} \sim 3000$--$4000\, \rm 
K$; these are  classified as S-type (stellar) systems. The remaining 20\,\% of symbiotic 
systems contain Mira-type variables and their near-IR  colours are consistent with the 
combination of a reddened Mira and warm ($T \sim 1000\, \rm K$) dust shell; these are  
classified as D-type (dusty)  systems. 

\begin{figure} \plotone{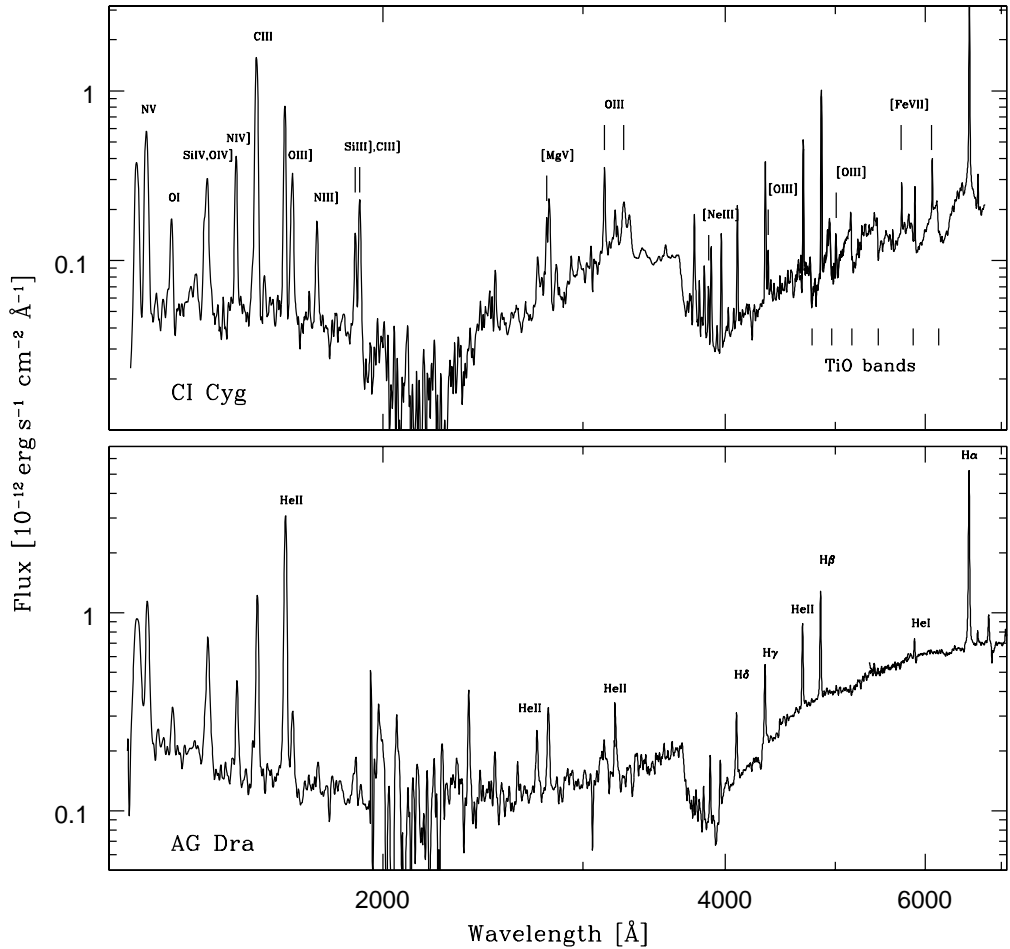} \caption{Quiescent optical/ultraviolet spectra for 
symbiotic stars. Strong TiO bands in the red spectrum of CI Cyg, indicate an M4$-$5\,III 
primary; they are absent in the yellow, galactic halo system AG Dra with a K3\,II--III 
primary. The intensity of the UV continuum and He\,{\sc ii} emission lines require similar 
temperature, $\ga 10^5$ K, and luminosity, $\sim 1000\, \rm L_{\sun}$, of the hot 
component in both systems, whereas the CNO emission lines in AG Dra are systematically 
weaker than those in CI Cyg which reflects low metallicity of AG Dra. Another difference 
is the presence of strong high-excitation forbidden lines (e.g. [Fe\,{\sc vii}], [Mg\,{\sc 
v}], [O\,{\sc iii}]) in CI Cyg which are not visible in AG Dra.} \end{figure}

Since the presence of an evolved red giant is indispensable to make a symbiotic binary, 
the system must have enough space for such a large star. Symbiotic stars are thus 
interacting binaries with the largest  orbital separations, and their study is essential 
to understand the interactions of detached and semi-detached binary stars. Mass accretion 
onto the hot component also plays a very important role in driving the basic properties 
and evolution of symbiotic stars, and involves energetic phenomena relevant to many other 
astrophysical objects. The presence of both an evolved giant with heavy mass loss and a 
hot companion copious in ionising photons also results in a rich and luminous 
circumstellar environment surrounding the interacting stars. 

Such a complex multi-component structure makes symbiotic stars a very attractive 
laboratory to study various aspects of stellar evolution in binary systems. It is worth to 
mention here the very interesting relatives of symbiotic stars: GX 1+4 -- consisting of  
an M6 III giant and a neutron star companion, and the peculiar black hole binary GRS 
1915+105 recently discovered to comprise a K giant companion (Greiner et al. 2001), which 
are among the most exotic and variable X-ray binaries.
\begin{figure} \plotone{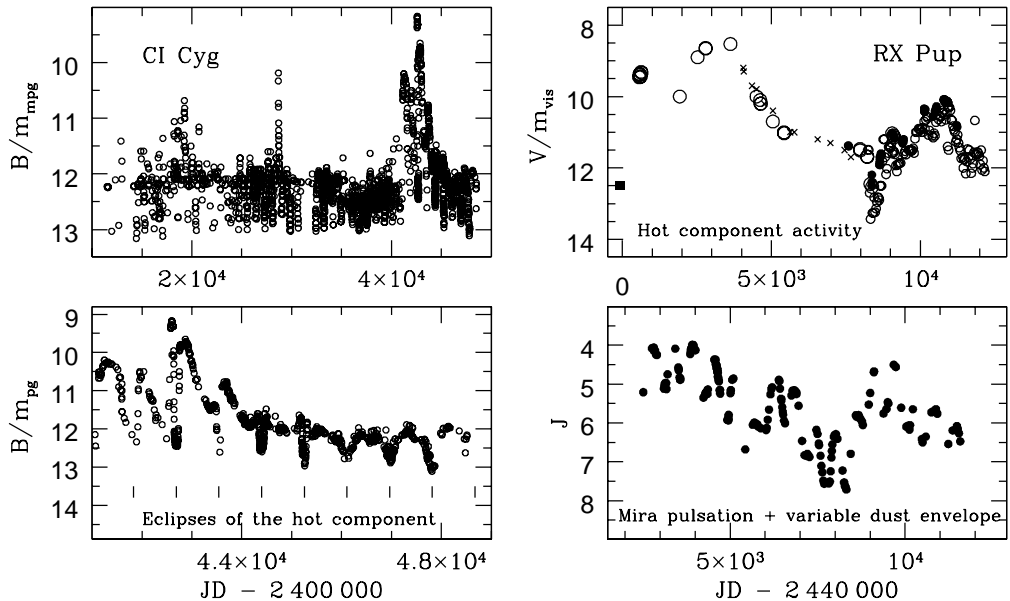} \caption{(left) The hot component outbursts (top) 
and deep eclipses (bottom) are the most prominent features of the 1894-1991 B/mpg light 
curve of the classical S-type symbiotic system CI Cyg. (right) The optical light curve 
(top) of the D-type system  RX Pup is dominated by the hot  component activity whereas the  
near-IR light (bottom) is dominated by the Mira pulsation and variable obscuration of the 
Mira by its circumstellar dust.} \end{figure} 
It is very important to set firm 
links between symbiotic stars and related objects, in order to understand the role of 
these binaries in the formation of stellar jets, planetary nebulae, novae, supersoft X-ray 
sources and SN Ia. Many of those are issues concerning the late stages of stellar 
evolution which are presently poorly known, but with  important implications on our 
understanding of stellar pulsations and chemical evolution of galaxies, as well as of the 
extragalactic distance scale.

The goals of this presentation is to present recent progress in the determination of 
orbital and stellar parameters of symbiotic stars, and to provide input for more detailed 
discussion of such important topics as: basic properties of the symbiotic cool giant 
components compared to single evolved giants; the nature of the hot component and the 
physics of its outbursts; the origin of the symbiotic circumstellar nebulae, and their 
links with planetary nebulae; evolutionary paths leading to symbiotic stars and to their 
progeny, and many others. We start with a brief description of variable phenomena found in 
symbiotic stars in Section 2. The recent progress in determination of their orbital and 
stellar parameters is summarized in Section 3. We discuss basic properties of the 
symbiotic giants compared to single evolved giants (note that symbiotic Miras are 
discussed in detail by P.A. Whitelock) in Section 4 as well as the nature of the hot 
component and its outburst evolution in Section 5. 

\section{Light curves and variability}

The composition of symbiotic stars, specifically the presence of a red giant and its 
accreting companion, places them among the most variable stars. 
They can fluctuate in many different ways. 
In addition to periodic changes due to the binary motion, both the cool giant and 
the hot component can also show intrinsic variability (Fig. 2). For the red 
giant,  these can be radial pulsations (all D-type and some S-type systems) and 
semiregular variations (S-type) of the giant, with timescales of the order of months and 
years as well as long-term light changes due to variable obscuration by circumstellar dust 
(mostly D-type), solar-type cycles, etc. Mass accretion onto the hot component also 
results  in different variable phenomena with timescales from seconds and minutes 
(flickering and QPO)  to years and decades (nova-like eruptions). The examples of light 
curves for well-studied, though not yet completely understood, systems: RX Pup, 
CI Cyg and CH Cyg, which are representative for different variabilities observed 
in these objects are also presented and discussed by Miko{\l}ajewska (2001). 

The wealth of variable phenomena found in symbiotic stars is really a challenge for 
patient observers in any spectral range and  makes these systems excellent targets for 
long-term monitoring programs, especially that they do not need large telescopes. This 
also explains how and why symbiotic systems could so effectively elude their physical 
nature. In fact, the binary nature of all symbiotic stars was definitely confirmed, by 
both direct measurement of the binary motion as well as direct detection of the hot 
component continuum in ultraviolet spectral range, only in the 1980's, whereas various 
problems concerning their intrinsic variability and its relation to their binary nature 
remain still unresolved.

\section{Orbital parameters}

Although periodic photometric changes, with periods of 200--1000 days and amplitudes 
increasing towards short wavelengths, were detected in many symbiotic stars already in the 
1930s, they were generally not associated with binary motion. The first symbiotic stars 
announced to be eclipsing binaries were AR Pav (Mayall 1937) and CI Cyg (Hoffleit 1968). 

By now, the orbital periods are known for about 40 systems (Belczy{\'n}ski et al. 2000; 
Table 1). About half of them are eclipsing binaries whereas the reflection effect is 
responsible for the periodic light modulation in other systems. So far, only six systems, 
T CrB, CI Cyg, EG And, BD--21\,3873, BF Cyg and YYHer (Belczy{\'n}ski et al. 2000; 
Miko{\l}ajewska et al. 2002a,b) show the ellipsoidal light variations of a tidally  
distorted red giant.
 \begin{table} \caption{Orbital elements for symbiotic binaries} \footnotesize 
\begin{tabular}{l c c c r c l c l l } \tableline 
Star & $P$ & $K_g$ & $q$ & $\gamma_0$~~ & 
$e$ & ~~T$_{0}$& $a_g \sin i$ & $f(M)$ & \\
 & [days]& [km/s] & $M_{\rm g}/M_{\rm h}$& 
[km/s] &  & [JD$^1$] & [R$_\odot$] & [M$_\odot$] & \\ \tableline 
EG And & 482.6 & 7.3 & &  -95.0 & 0& 50804$^3$ & ~70 & 0.020 &[1]\\ 
AX Per & 682.1 & 7.8 & 2.3 & -117.4 & 0 & 50964& 105 & 0.033 & [2]\\ 
BD Cam & 596.2 & 8.5& & -22.3& ~0.09& 42794$^2$& ~99.7& 0.037& [1] \\ 
V1261 Ori & 642& 7.5& & 79.7& 0.07& 46778$^3$& 95& 0.028& [1] \\ 
BX Mon & 1401 & 4.3 & 6.7 & 29.1 & 0.49 & 49530& 104 & 0.0076 & [1] \\
 & 1259 & 4.6 &     & 29.1 & 0.44 & 49680& 103 & 0.0092 & [1] \\ 
SY Mus & 624.5 & 7.4 & & 12.9 & 0 & 49082$^3$& ~91 & 0.026& [1] \\ 
TX CVn & 199 & 5.7 & & 2.3 & 0.16 & 45195$^2$& ~22 & 0.004 & [1] \\ 
RW Hya & 370.2 & 8.8 & &12.4 & 0 & 45072 & ~65 & 0.026 & [1] \\
 & 370.4 & 8.8 & & 12.9 & 0 & 49512&     & 0.026 & [1] \\ 
BD-21\,3873 & 281.6 & 10.6& & 203.9& 0& 49087$^3$& ~59& 0.035& [1] \\
 T CrB & 227.57 & 23.9 & 0.6 & -27.8 & 0 & 47919$^3$& 107 & 0.322 & [1] \\ 
AG Dra & 549 & 5.9 & & -147.2 & 0 & 50775 & ~64 & 0.0115& [2] \\ 
KX TrA & 1350 & 6.8 & 2.3 & -123.7 &  0.29 & 51703 & 175 & 0.039 & [4] \\ 
AE Ara & 812 & 5.4 & 4 & -15.7 & 0 & 50217 & 87 & 0.0133 & [5] \\ 
RS Oph & 455.7 & 16.7 & 0.35 & -40.2 & 0 & 50154$^3$ & 150 & 0.221 & [1]\\ 
V343 Ser & 451.3 & 2.6 &  & -5.63 & 0 & 50398$^3$ & 23 & 0.0008 & [3]\\
 & 450.5 & 2.7 &  & -5.65 & 0.14 & 50575$^2$ & 23.5 & 0.0009& [3]\\ 
FG Ser & 633.5 & 6.9 &  & 73.3 & 0 & 51031 & 87 & 0.022 & [2]\\ 
AR Pav & 604.5 & 10.9& 2.5 & -68.3& 0 & 48139& 130& 0.079 & [6] \\ 
V443 Her & 594 & 2.5 & & -55.5 & 0 & 50197 & ~30 & 0.0010 & [2] \\
FN Sgr & 568.3 & 10.5 & 2.1 & -53.7 & 0 & 50269 & 118 & 0.0689 & [7]\\ 
BF Cyg & 757.2 & 6.7 & 3.6 & -3.75 & 0 & 51395 & 100 & 0.0239 & [3]\\ 
CH Cyg & 5700 & 4.9 & & -57.7 & 0.47 & 45086& 478 & 0.045 & [1] \\
 & 756.0 & 2.6 & & -60.6 & 0 & 46644& ~39 & 0.0014 & [1] \\
 & 5292 & 4.8 & &        & 0.06 & 45592$^2$& 500 & 0.060 & [1] \\ 
CI Cyg & 855.3 & 6.7 & 3 & 18.4 & 0 & 45242& 114 & 0.027 & [1] \\
 &  853.8 &  6.7 &   & 15.0 & 0.11 & 50426& 112 & 0.026 & [1] \\ 
V1329 Cyg & 956.5 & 7.9 & 2.9 & -23.1 & 0 & 51565 & 149 & 0.0481 & [3]\\ 
CD-43\,14304 & 1448& 4.4& & 27.6& 0& 45929$^3$& 126 & 0.013& [1] \\
 & 1442& 4.6& & 27.5& 0.22& 45560$^2$& 128& 0.014& [1] \\ 
AG Peg & 816.5 & 5.3 & 4 & -15.9 & 0 & 31668& ~84 & 0.012 & [1] \\
 & 818.2 & 5.4 &   & -15.9 & 0.11 & 46812& ~87 & 0.0135 & [1]\\ 
Z And & 758.8 & 6.7 &  & -1.8 & 0 & 50260& 102 & 0.024& [2] \\ 
CD-27\,8661& 763.3& 10.5& & -5.5& 0& 49280$^3$& 158 & 0.092& [1]\cr 
\tableline \tableline \end{tabular} 

\noindent T$_{0}$ -- time of inferior spectroscopic conjunction of the giant; $^1$ -- 
Julian Date = 2\,400\,000 + JD listed in table; $^2$ -- time of the passage through 
periastron; $^3$ time of maximum velocity.\\ References: [1] -- Belczy{\'n}ski et al. 
2000, and references therein; [2] -- Fekel et al. 2000, and references therein; [3] -- 
Fekel et al. 2001, and references therein; [4] -- Ferrer et al. 2002; [5] -- 
Miko{\l}ajewska et al. 2002e; [6] -- Quiroga et al. 2002a,b; [7] -- Brandi et al. 2002. 
\end{table} 
 All systems with measured orbital period but R Aqr belong to the S- 
type, and most of them have periods between $\sim 200$--1000 days (Table 1; Fig. 3). 
Moreover, all systems with $P_{\rm orb} \ga 1000$ days except for the yellow symbiotic CD--
 43\,14302 contain very cool giants with spectral type $\ga \rm M6$. The longest orbital 
period  thus far estimated in symbiotic system is the 44-yr period of the symbiotic Mira R 
Aqr. The orbital periods of symbiotic Miras with thick dust envelopes (D-type) could be 
even larger. In general, the binary separations in D-type systems must be larger than the 
dust formation radius. Assuming  a typical  dust formation radius of $\ga 5 \times R_{\rm 
Mira}$, and $R_{\rm Mira} \sim 1$--$3\, \rm au$, the minimum binary separation is $a \ga 
20\, \rm au$, and the corresponding binary period is $P_{\rm orb} \ga 50\, \rm  yr$, for 
{\it any} D-type system. Thus the orbital period distribution seems to be the result of 
compromise between the minimum binary separation yet providing enough space for the 
evolved giant and the minimum mass accretion rate required for triggering the symbiotic 
phenomenon.

\begin{table} 
\caption{Mass estimates for symbiotic binaries}

\footnotesize

\begin{tabular}{l c c c  c  c  l } 
\tableline 
Star & $P$\,[days] & Ecl. & $i$\,[deg] & $M_{\rm g}[\rm M_\odot]$ & $M_{\rm h}[\rm 
M_\odot]$ & Com. \\ 
\tableline 

EG And & 481 & Y & 90 & $1.5\pm0.6$ & $0.4\pm0.1$ & ET \cr
AX Per & 680.8 & Y & 90 &  $0.9\pm0.2$ & $0.37\pm0.06$ & BA \\
             &          &    & $\ga 70$  & $\la 1.1$  & $\la 0.44$ \cr
BX Mon & 1401 &  Y & 90 & $3.0\pm1.5$ &  $0.45\pm0.21$ &  BA \\
              &         &      & $\ga 62$ & $\la 3.7$ & $\la 0.6$ \cr 
SY Mus & 625 & Y & 90 & $1.3 \pm 0.25$ & $0.43 \pm 0.05$ & ET \cr 
RW Hya & 370.2 & Y & 90 & $1.6\pm0.3$ & $0.48\pm0.06$ & ET \cr
T CrB & 227.57 &  N & $\sim 60$ & $0.7\pm0.2$ & $1.2\pm0.2$ &  BM \\
KX TrA & 1350 & ? &  90 & $1.0\pm0.3$ & $0.41\pm0.04$ & He\,{\sc ii} W\\
              &         &    & 135 & 2.7 & 1.2 & SP \\
AE Ara & 812 &  N & 60 & $2.0\pm1.2$ & $0.51\pm0.2$ & He\,{\sc ii}  W\\
RS Oph & 455.7 & N &  $\leq 45$ & $\geq 0.40$ & $\geq 1.1$ & BA. \\ 
FG Ser & 650 & Y & 90 &$1.7\pm0.7$ & $0.60\pm0.15$ & ET  \cr
AR Pav & 604.5 & Y & 90 & $2.5\pm0.6$ & $1.0\pm0.2$ & BA \\
             &          &     & $\ga 70$ &  $\la 3$ & $\la 1.2$ & \\
&  & Y & 90 & $2.0\pm0.5$ & $0.87\pm0.15$ & ET \cr
FN Sgr & 568.3 & Y & 90 & $1.4\pm0.2$ & $0.66\pm0.08$ & BA. \\
            &          &     & $\ga 70$  & $\la 1.7$  &  $\la 0.8$ \cr
BF Cyg & 757.2 & Y & 90 & $1.8\pm0.6$ & $0.51\pm0.1$ & UVEL \cr
             &          &     & $\ga 70$ & $\la 2.2$ & $\la 0.6$ \cr
CI Cyg & 855.3 & Y & 90 & $1.3\pm0.3$ & $0.43\pm0.04$ & He\,{\sc ii} EL \\
            &          &     & $\geq 79$ & $\leq 1.6$ & $\leq 0.52$ \cr
V1329 Cyg & 956.5 & Y & 86 & $2.1\pm0.5$ & $0.74\pm0.08$ & H\,{\sc i} W, SP \\
AG Peg & 816.5 & N & $\la 60$ & $\ga 1.8$ & $\ga 0.46$ & He\,{\sc ii} EL \\
\tableline
\tableline 
\end{tabular} 

\noindent BA -- blue absorption system; UVEL -- ultraviolet emission lines; He\,{\sc ii} W 
-- He\,{\sc ii} emission wings; H\,{\sc i} W -- H\,{\sc i} emission wings;  He\,{\sc ii} 
EL -- He\,{\sc ii} emission line; SP -- $i$ from spectropolarimetry; ET -- the cool giant 
mass from $v\,\sin i$ and evolutionary tracks (M{\"u}rset et al. 2000, and references 
therein; Schild et al. 2001); BM -- light curve synthesis combined with radial velocity 
curve (Belczy{\'n}ski \& Miko{\l}ajewska 1998). \noindent \end{table}

In a binary with $P_{\rm orb} \ga 1\, \rm yr$, the expected amplitude of radial velocity 
changes is rather low, $K_{\rm g} \sim 5$--$10\, \rm km\,s^{-1}$. Nevertheless, modern 
observations with photon-counting detectors and new methods of analysis have allowed to 
derive spectroscopic orbits from the radial velocities of the red giant for most of the 
bright symbiotic stars (Table 1). It is interesting that vast majority of symbiotic stars 
have circular (or nearly circular, $e \la 0.1$) orbits, and a significant eccentricity has 
been found only for BX Mon, KX TrA, CH Cyg and CD$-$43\,14304, which have also the 
longest orbital periods, $P_{\rm orb} > 1000$ days. The eccentricity-period distribution 
for symbiotic stars is significantly different from the $e$-$P_{\rm orb}$ distribution in 
other binaries with late-type giants, which show significant fraction of eccentric orbits 
among systems with $P_{\rm orb} \la 1000$ days (e.g. Fig. 9 of Jorissen \& Mayor 1992). 

\begin{figure} \plotone{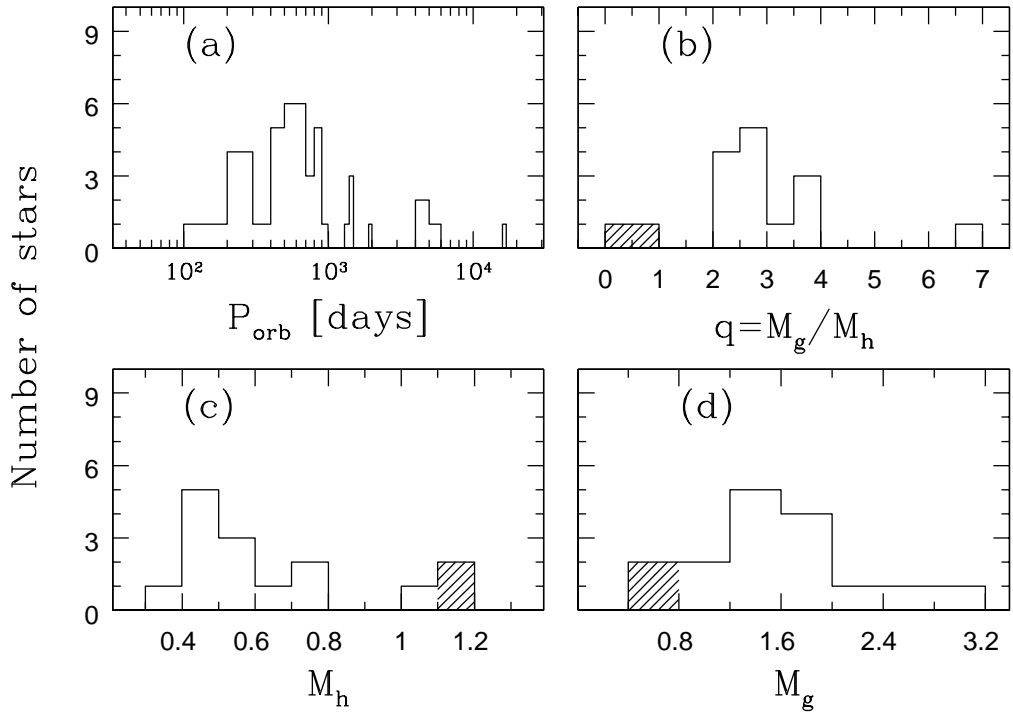} \caption{Distribution of (a) binary periods, (b) 
mass ratios, (c) hot component and (d) cool giant masses, respectively. The shaded regions 
denote the symbiotic recurrent novae.} \end{figure}

In several cases, the motion of the hot component was also measured from either the blue 
absorption features or the optical and UV emission lines and broad emission line wings, 
and the masses of both components could be derived (Table 2). For some systems, the 
component masses were derived by combining the mass function for the red giant with the 
giant mass calculated from $v \sin i$ measurements and evolutionary tracks (e.g. 
M{\"u}rset et al. 2000). The distributions of mass ratios and masses of the symbiotic 
binary components are shown in Figure 3b$-$d. All systems except BX Mon and the symbiotic 
recurrent novae T CrB and RS Oph have the mass ratio, $q=M_{\rm g}/M_{\rm h}$, between 2 
and 4. The very high mass ratio for BX Mon is based on only 2 measurements of the radial 
velocities of the blue absorption lines, and has to be confirmed by further observations 
covering at least one orbital period. The cool giant masses are all between 0.6--$3.2\, 
\rm M_{\rm sun}$, with a peak around $1.6\, \rm M_{\sun}$ (the average is $1.7\pm0.1\, \rm 
M_{\sun}$). The masses of the hot component are all between 0.4--$0.8\, \rm M_{\sun}$ 
(with the average at $0.53\pm0.05\, \rm M_{\sun}$) except those of AR Pav,  T CrB, and RS 
Oph, and they fall into the range of masses expected for white dwarfs. 
In both symbiotic recurrent novae, the cool giant is the less 
massive component, and its mass, $M_{\rm g} \la 0.8\, \rm M_{\sun}$ is lower than the mass 
of any other symbiotic giant, whereas their hot components, 
with $M_{\rm h} \sim 1.1$--$1.3\, \rm M_{\sun}$ are the most massive, and their masses are 
sufficient to become a Ia supernova. 

Finally, the direct measurements of hot component masses can be used to test 
the possible relation between white dwarf mass and orbital period (Miko\-{\l}a\-jew\-ska 
\& 
Kenyon 1992a). The data from Table 2 are plotted in Figure 5 (left panel), and they do 
not 
show any significant correlation with the orbital period. We must, however, note that the 
estimates for BX Mon and KX Tra, the only systems with longer periods, $P_{\rm orb} > 
1000^{\rm d}$, are among the most uncertain results in Table 2.

\section{The cool giant}

Based on near IR spectra, M{\"u}rset \& Schmid (1999) classified the cool giants in about 
100 systems, and found that for S-type systems the spectral types cluster strongly between 
M3 and M6, with a peak at M5, whereas the distribution of symbiotic miras peaks at 
spectral types M6 and M7. They also noticed a strong bias towards later spectral types 
when compared to  red giants in solar neighborhood, and that the frequency of mira 
variables is higher among symbiotic giants. This predominance of very late, and thus more 
evolved,  giants in symbiotic systems indicates that large radius and high mass 
loss from the cool giant is the key parameter for triggering the symbiotic phenomenon 
in binaries. 

The process of mass transfer -- Roche lobe overflow or stellar wind -- is one of 
fundamental question in relation to symbiotic binaries. Although, wind accretion is 
obviously the case for companions of symbiotic Miras, at least in some of the S-type 
systems, the cool giant could, in principle, fill its tidal lobe. Amongst  the main 
evidence for the predominance of wind-accretion is the fact that ellipsoidal light 
variations, characteristic of tidally distorted stars, are rarely observed for symbiotic 
stars. However, the general absence of such changes may be in fact 
due to the general lack of systematic searches for the ellipsoidal variations in the red 
and near-IR range where the cool giant dominates the continuum. This is best illustrated 
by CI Cyg, BF Cyg and YY Her, in which only the 
quiescent $VRI$ and near-IR light curves show a modulation with half-orbital period as 
expected for ellipsoidal variability of the red giant whereas the $UB$ light curves do not 
show clear evidence for such changes (Miko{\l}ajewska 2001; Miko{\l}ajewska et al. 
2002a,b). Interestingly, such changes have been thus far found only in systems with 
multiple outburst activity whereas there are apparently absent in two non-eruptive 
systems, V433 Her and RW Hya (Miko{\l}ajewska et al. 2002c).

Based on a sample of 8 symbiotic systems with accurate orbits and estimated radii of the 
red giants, Schild et al. (2001) argued that none of the symbiotic giants with known radii 
(derived from the measured $v\sin i$ in 6 cases, and from eclipse contacts in AX Per) 
fills its Roche lobe. We note, however, that they did not included in their sample any of 
the systems  with evident ellipsoidal changes. Similarly, Fekel, Hinkle \& Joyce (2002) 
derived the $v \sin i$ values and estimated the cool giant sizes for 13 symbiotic systems. 
Their sample includes CI Cyg, BF Cyg and T CrB, however only CI Cyg has  large Roche lobe 
filling factors 
$R_{\rm g}/R_{\rm L}\sim 0.8$, whereas the giants in both BF Cyg and T 
CrB, with  $R_{\rm g}/R_{\rm L}\sim 0.3$ and $\sim 0.4$, are far from filling their tidal 
lobes. The rotational velocity consistent with a synchronously rotating Roche lobe-filling 
giant was found in  CI Cyg and AX Per (Kenyon et al. 1991; Miko{\l}ajewska \& Kenyon 
1992b). Recently, Orosz \& Hauschildt (2000) have shown that rotational broadening kernels 
for tidally distorted giants can be significantly different from 
analytic kernels due to a combination of the nonspherical shape of the giant and the 
radical departure from a simple limb darkening law. As a result, geometrical information 
inferred from $v \sin i$ measurements of cool giants in binary systems, and in particular 
in symbiotic stars, is likely biased. Summarizing, in our opinion, the question whether 
the symbiotic giants, at least in some of the active S-type systems, are tidally distorted 
or not remains open as long as we do not have good near-IR light curves which would 
confirm or exclude such possibility. 

Mass-loss rates for the symbiotic giants 
derived from either the cm and mm/submm radio observations (Seaquist, Krogulec \& Taylor 
1993; Miko{\l}ajewska, Ivison \& Omont 2002a,b) or from analysis of IRAS data (Kenyon, 
Fernandez-Castro \& Stencel 1988) are of order of $10^{-7}\, \rm M_{\sun}\,yr^{-1}$, and 
they are systematically higher than those reported for single M giants, which again 
suggests that high mass-loss rate for the giant is essential to produce the symbiotic 
star. 

Whitelock \& Munari (1992) suggested that the symbiotic giants may be related to the 
metal-rich M stars found in the Galactic Buldge and elsewhere, i.e. they have low masses,  
$\sim 1\, \rm M_{\sun}$, and $Z \ga Z_{\sun}$. They also noted  that the mass-loss rates 
of the symbiotic giants although systematically greater than for the local bright giants 
are similar to those of the Bulge-like stars. Their findings, however, have not been 
confirmed by direct estimates of elemental abundances. In particular, Nussbaumer et. al. 
(1988) found that the CNO abundance ratios deduced from UV emission lines for 24 symbiotic 
stars are best fitted by normal red giants. Moreover, carbon abundances and $\rm 
^{12}C/^{13}C$ ratios have been obtained by fitting  synthetic spectra to the observed  
first-overtone CO absorption features in K-band spectra of 7 northern and 6 southern 
symbiotic systems (Schmidt \& Miko{\l}ajewska 2002, and references therein). In all cases, 
subsolar carbon abundances and $\rm ^{12}C/^{13}C$ have been found, which indicates that 
the surveyed symbiotic giants are indistinguishable from local normal M giants, in 
agreement with the abundance studies based on nebular emission lines. 

\section{The hot component and activity}

The typical hot components of symbiotic binaries appear to be quite hot ($T_{\rm eff} \ga 
10^5\, \rm K$) and luminous ($L_{\rm h} \ga 10^2 - 10^3\, \rm L_{\sun}$) and overlap into 
the same region in the H-R diagram as the central stars of planetary nebulae (Fig. 4; 
M{\"u}rset et al. 1991; Miko{\l}ajewska, Acker \& Stenholm 1997). There is also no 
significant difference between the symbiotic hot components in our Galaxy and those in the 
Magellanic Clouds, except that the latter might be slightly hotter. It is hard to believe 
that such high luminosities are powered solely by accretion, as they would require 
accretion rates of $\sim\, \rm a\,few \times 10^{-6} M_{\sun}\,yr^{-1}$. Although 
symbiotic giants have high mass loss rates, of $\sim 10^{-7}\, \rm M_{\sun}\,yr{- 1}$ and 
more, in most cases they interact via stellar wind instead of Roche lobe overflow, and the 
expected accretion rate is an order of magnitude lower. The situation radically changes
 if the symbiotic white dwarfs burn hydrogen-rich material 
as accrete it. The accretion rate of order of $10^{-8}\, \rm M_{\sun}\,yr^{-1}$ is then 
sufficient to power the typical hot component with $M_{\rm h} \sim 0.5\, \rm 
M_{\sun}\,yr^{-1}$. 

According to the theoretical studies, the hydrogen-shell burning luminosity should be a 
function of the underlying core mass. Unfortunately, there is no unique core 
mass-luminosity relationship for accreting white dwarfs because it depends significantly 
on the 
thermal history of the white dwarf: generally hot white dwarfs must have larger masses 
than the cold ones to reach the same luminosity during the hydrogen-burning phase (Iben \& 
Tutukov 1996). However, 
\begin{figure} \plotone{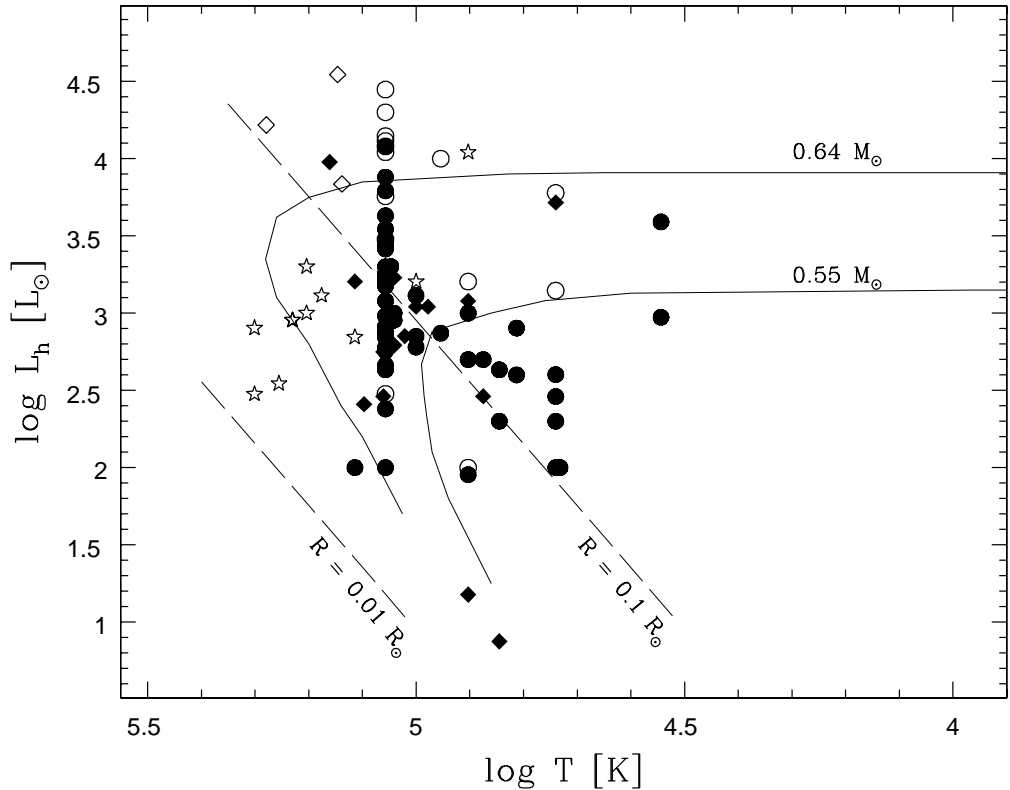} \caption{The hot components of symbiotic stars in the 
H-R diagram. Data from M{\"u}rset et al. (1991) and Miko{\l}ajewska, Acker \& Stenholm 
(1997) are plotted as diamonds and circles, whereas filled and open symbols represent the 
galactic S- and D-types, respectively. Stars correspond to the symbiotic stars in the 
Magellanic Clouds (Morgan 1992; 1996). The solid curves are the evolutionary tracks from 
Sch{\"o}nberner (1989), the dashed lines correspond to constant radii.} \end{figure} 
for the symbiotic systems with known component masses (Table 2) we 
are in a position to determine which of the theoretical mass-luminosity relations is 
applicable in their case. The positions of the symbiotic hot components in the luminosity 
versus mass plane together with different mass-luminosity relationships are plotted in 
Figure 5 (right panel). 
The symbiotic white dwarfs seem to cluster around the mass-luminosity relations for stars 
leaving the AGB with CO core, and those leaving RGB with a 
degenerate He core, for the first time. It is possible that the white dwarf descendant of 
the more massive component could still be hot at the onset of mass transfer from the less 
massive red giant.

The hot components in many symbiotic systems show intrinsic variability (Fig. 2). Based on 
this activity we distinguish between ordinary or classical symbiotic stars, which show 
occasionally 1$-$3 mag eruptions with time scales from months to a few years (such as Z 
And, CI Cyg and AG Dra), and symbiotic novae that have undergone a single outburst of 
several magnitudes lasting for dozen of years. Figure 6 shows outburst evolution of 
selected symbiotic systems. The outburst behaviour of AG Peg and RX Pup is consistent with 
a thermonuclear nova eruption. In both systems the outburst develops very slowly: the rise 
to maximum takes months, and  the decline to the pre-outburst stage lasts dozens of years. 
AG Peg is the record-holder among all symbiotic novae: its eruption began in 1850's, and 
the hot component maintained a constant luminosity, $L_{\rm h} \sim 3000\, \rm L_{\sun}$, 
for at least 100 years. The evolution of RX Pup was much faster, the constant luminosity 
(plateau) phase lasted for only 11 years, and the maximum plateau luminosity, $<L_{\rm 
h}> \sim 15\,000\, \rm L_{\sun}$ was about 5 times higher than that of AG Peg. These 
difference in the outburst evolution can be accounted for by different white dwarf masses: 
higher in RX Pup than that in AG Peg. RX Pup is also a possible recurrent nova.

\begin{figure} \plotone{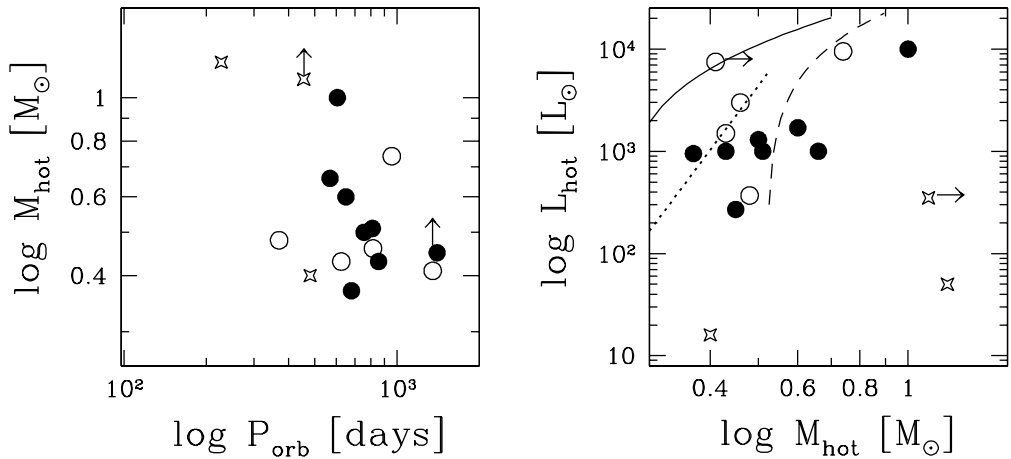} \caption{Mass-orbital period relation (left), 
and the mass-luminosity relation (right) for the hot components. Closed symbols represent 
systems with Z And-type multiple outburst activity, open circles -- noneruptive systems 
and symbiotic novae, and stars -- accretion-powered systems, namely, symbiotic recurrent 
novae at quiescence and EG And. The upper solid curve is mass-plateau luminosity 
relationship for accreting cold white dwarfs (Eq. 6 of Iben \& Tutukov 1996),  the dashed 
curve is the Paczy{\'n}ski-Uus relation (e.g. Paczy{\'n}ski 1970) for AGB stars with a CO 
core, and the dotted one the relation for helium dwarfs (Eg. 7 of Iben \& Tutukov 1996).} 
\end{figure}

Unfortunately, the thermonuclear models do not account for the multiple outburst activity 
of Z And, AG Dra (Fig. 6), and other classical systems. In most of them, the hot  
component maintains a roughly constant luminosity whereas its effective temperature varies 
from $\sim 10^5$ to $\sim 10^4\, \rm K$, and Z And is a good example  of such an outburst. 
Contrary to this behaviour, luminosity of the hot component of AG Dra increases by a 
factor of $\sim 10$ whereas its temperature increases during the early stages of each 
eruption and then decreases as the outburst continues. A possible and promising 
explanation of this activity involves changes in mass transfer and/or accretion disk 
instabilities. There are many arguments in favour of such explanation. First, the hot 
companion of all symbiotic giants which fill or nearly fill their tidal lobes show the 
multiple outburst activity. Moreover, even if the giant does not fill its Roche lobe, its 
wind is likely focused towards the secondary and/or towards the orbital plane which would 
facilitate an accretion disk formation (e.g. Gawryszczak, Miko{\l}ajewska \& 
R{\'o}{\.z}yczka 2002; Mastrodemos \& Morris 1999). 

\begin{figure} 
\plotone{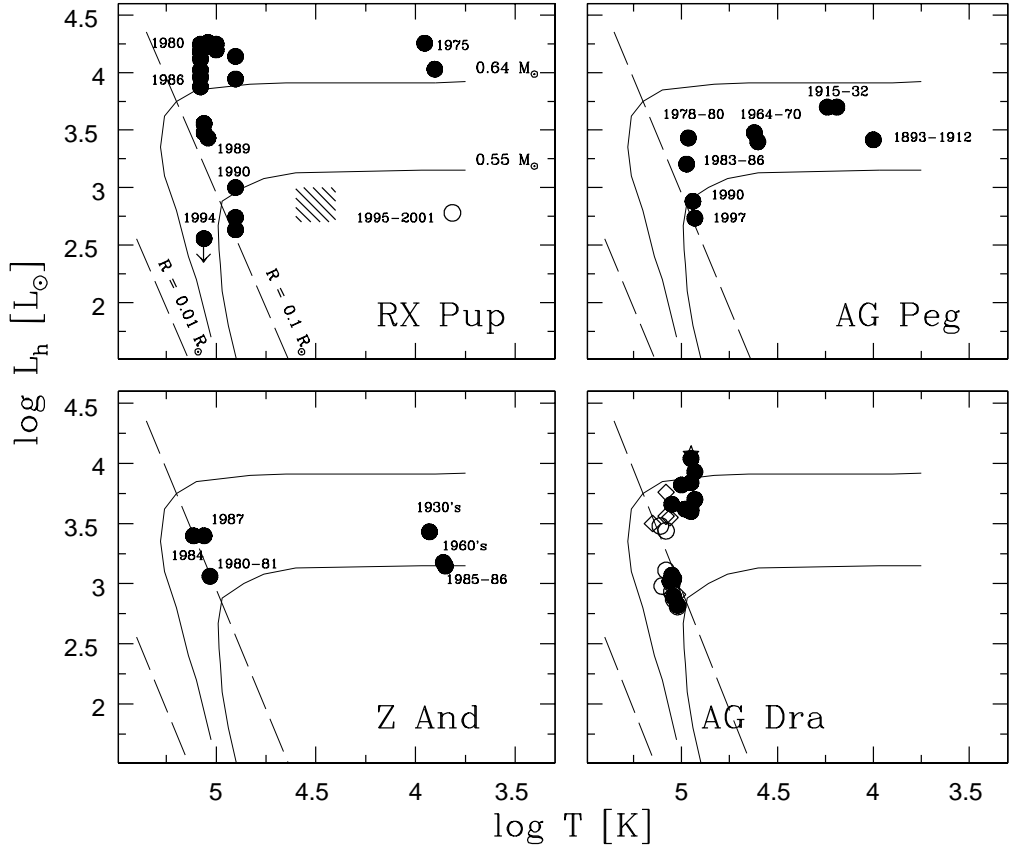} \caption{Outburst evolution of the hot component of the 
symbiotic novae RX Pup and AG Peg, Z And, and the yellow symbiotic AG Dra (Miko{\l}ajewska 
et al. 1999; Kenyon et al. 1993; Kenyon, Proga \& Keyes 2001; Miko{\l}ajewska \& Kenyon 
1996; Miko{\l}ajewska et al. 1995.)} \end{figure} 

The timescales for these eruptions are 
very similar to those of the hot component luminosity changes (high and low states) in 
symbiotic recurrent novae (T CrB, RS Oph and RX Pup) between their nova eruptions and the 
accretion-powered system of CH Cyg and MWC\,560. During bright phases, in all these 
systems the optical data indicate the presence of relatively cool B/A/F-type shell source 
while the UV/optical emission lines require a much hotter source with a roughly comparable 
luminosity (Miko{\l}ajewska et al. 2002d). Similar double-temperature structure of the hot 
component is observed in the Z And- type systems during their outbursts. In  AX Per, AR 
Pav, FN Sgr and possibly other active systems,  the A/F-type absorption lines trace the 
orbit of the hot component, and they are probably formed in a geometrically and optically 
thick accretion disc and in a gas stream (Miko{\l}ajewska \& Kenyon 1992b; Quiroga et al. 
2002a,b; Brandi et al. 2002). Other evidence for the presence of an accretion disks in the 
active symbiotic stars include: the ellipsoidal shape of the blue continuum source during 
outburst; presence of secondary periods (always 10--20\,\% shorter that $P_{\rm orb}$) in 
the outburst light curves, similar to the superhumps in CVs (Miko{\l}ajewska et al. 
2002c); bipolar geometry of outflows associated with the hot component activity in CI Cyg, 
AG Dra and other systems (Miko{\l}ajewska \& Ivison 2001; Miko{\l}ajewska 2002; Tomov, 
Munari \& Maresse 2000). It is possible that both the activity of the classical Z And-type 
symbiotic stars and the high and low states of accretion-powered systems are related to 
the presence of an unstable accretion discs, with the only difference that the hot 
component in the former burns more or less stably the accreted hydrogen whereas not  in 
the latter.

Finally, Miko{\l}ajewska \& Kenyon (1992a) argued that the outburst evolution of CI Cyg 
and AX Per are best explained by the presence of an unstable thick disk around a low-mass 
main-sequence accretor. We must, however, note that the quiescent characteristics of their 
hot components are consistent with a hot and luminous stellar source powered by 
thermonuclear burning. In particular, they fall in the same region in the HR diagram as 
the hot components of other symbiotic stars (Fig. 4). Their outburst evolution is also 
very similar to that of Z And, FN Sgr, and other multiple-outburst systems, which 
apparently points to common mechanism of the outburst and the nature of their hot 
component. 
Unfortunately, it will remain uncertain as long as we do not have a good theoretical model 
accounting for both their quiescent and outburst characteristics.

\section{Concluding remarks}

The nature of the cool giant plays the key role in the symbiotic phenomenon because it 
constrains the size of the binary system, which must have enough room for a red giant, and 
yet to transfer enough material to the companion and give rise to the symbiotic 
apparition. As a result, we have two distinct classes: the S-type with normal giants and 
orbital periods of $\la 15$ yr, and the D-type with Mira primaries and periods of 
$\ga 50$ yr.

Most S-type systems have $P_{\rm orb} \sim 200$--$900$ days and circular orbits.
A typical S-type system consists of a low-mass M3--6 giant ($<M_{\rm g}> \sim 1.7\, \rm 
M_{\sun}$)  transferring material to a hot, $\sim 10^5\, \rm K$, white  dwarf companion 
($<M_{\rm h}> \sim 0.5\, \rm M_{\sun}$). Although most symbiotic stars seem to interact by 
wind-driven mass loss, at least  some of the systems with multiple outburst activity may 
contain a Roche-lobe filling (or nearly filling) giants. 

Symbiotic M giants have subsolar carbon and $^{12}$C/$^{13}$C abundances, and they are 
indistinguishable in this respect from local M giants. They  have, however, systematically 
higher mass-loss rates than single giants, which suggests that high mass-loss rate for 
the giant is essential for triggering the symbiotic activity. The chemical abundances 
derived from nebular emission lines suggest that 
the main body of symbiotic nebulae is formed from material lost in the giant wind, 
whereas the hot companion is responsible for its ionization and excitation.

The typical hot component of a symbiotic binary appears to be a luminous, 
$\sim 100$--$10^4\, \rm L_{\sun}$, and hot, $\sim 10^5$ K, white dwarf powered by 
thermonuclear burning of the accreted hydrogen. 
The position of the symbiotic white dwarfs in the mass-luminosity plane indicates that 
they could be hot and luminous at the onset of mass transfer and symbiotic activity.
Although we still do not have satisfactory explanation of the multiple outburst activity,
there is increasing evidence that it may related to the presence of unstable accretion 
discs.

\acknowledgments This research was partly founded by KBN Research Grant No. 
5\,P03D\,019\,20.

\end{document}